\documentclass[aps,pra,twocolumn,showpacs,superscriptaddress,floatfix,10pt]{revtex4}
\usepackage[pdftex]{graphicx,color}
\usepackage{dcolumn}
\usepackage{bm}

\usepackage{verbatim}
\usepackage{amsfonts}
\usepackage{amssymb}
\usepackage{amsmath}
\usepackage{amsthm}

\usepackage[caption=false]{subfig}
\usepackage{rotate}
\usepackage{dsfont}

\let\oldmarginpar\marginpar
\renewcommand\marginpar[1]{\-\oldmarginpar[\raggedleft\tiny #1]%
{\raggedright\tiny #1}}

\def\KeyWord#1{$\backslash$\IfColor{$\!\!$\textRed{#1}\textBlack}{#1}$\!\!$}

\newcommand{\be}{\begin{equation} }
\newcommand{\ee}{\end{equation} }
\newcommand{\ba}{\begin{eqnarray} }
\newcommand{\ea}{\end{eqnarray} }

\newcommand{\bit}{\begin{itemize}}
\newcommand{\eit}{\end{itemize}}

\newcommand{\J}{J}
\newcommand{\AJ}{A_{\J}}
\newcommand{\BJ}{\bar{\J}}
\newcommand{\h}{\Gamma}
\newcommand{\Ah}{A_{\h}}
\newcommand{\Bh}{\bar{\h}}
\newcommand{\Q}{Q}

\def\cexp#1{\left[#1\right]}
\def\fracp#1{\left\|#1\right\|}

\newcommand{\typ}{\mathrm{typ}}

\graphicspath{{../FinalFigures/}}

\begin{document}

\title{The quasi-periodic quantum Ising transition in 1D}

\author{P.J.D. Crowley}
\email{philip.jd.crowley@gmail.com}
\affiliation{Department of Physics, Boston University, Boston, MA 02215, USA}

\author{A. Chandran}
\affiliation{Department of Physics, Boston University, Boston, MA 02215, USA}

\author{C.R. Laumann}
\affiliation{Department of Physics, Boston University, Boston, MA 02215, USA}

\date{\today}

\begin{abstract}
Unlike random potentials, quasi-periodic modulation can induce localisation-delocalisation transitions in one dimension. 
In this article, we analyse the implications of this for symmetry breaking in the quasi-periodically modulated quantum Ising chain. 
Although weak modulation is irrelevant, strong modulation induces new ferromagnetic and paramagnetic phases which are fully localised and gapless. 
The quasi-periodic potential and localised excitations lead to quantum criticality that is intermediate to that of the clean and randomly disordered models with exponents of $\nu=1^{+}$, and $z\approx 1.9$, $\Delta_\sigma \approx 0.16$, $\Delta_\gamma\approx 0.63$ (up to logarithmic corrections). 
Technically, the clean Ising transition is destabilized by logarithmic wandering of the local reduced couplings.
We conjecture that the wandering coefficient $w$ controls the universality class of the quasi-periodic transition and show its stability to smooth perturbations that preserve the quasi-periodic structure of the model. 
\end{abstract}

\maketitle

Sufficiently strong quasi-periodic modulation can drive a localisation transition in one dimensional wires, as was first shown by Azbel, Aubry and Andr\'e \cite{Aubry:1980aa,Azbel:1979aa,Hofstadter:1976aa,Harper:1955kl}. 
Insofar as the unmodulated wire is described by a critical Dirac theory, this suggests that strong modulation ought to be able to localise other quantum critical points.
On the other hand, if the critical point mediates the development of long-range order, it must have an extended mode at zero energy.
At the quantum Ising transition in the presence of \emph{disorder}, this tension gives rise to infinite randomness physics~\cite{McCoy:1968aa,mccoy1969theory,ma1979random,Shankar:1987aa,fisher1992random,fisher1995critical,fisher1999phase,motrunich2000infinite,igloi2005strong}. 
In this article, we show that sufficiently strong smooth \emph{quasi-periodic} modulation drives the quantum Ising transition into a new quasi-periodic (QP) Ising universality class. The properties of this universality class are intermediate between those of the clean and infinite randomness cases, and are robust to smooth perturbations that preserve the QP structure.

\begin{figure}
\begin{center}
\includegraphics[width=\columnwidth]{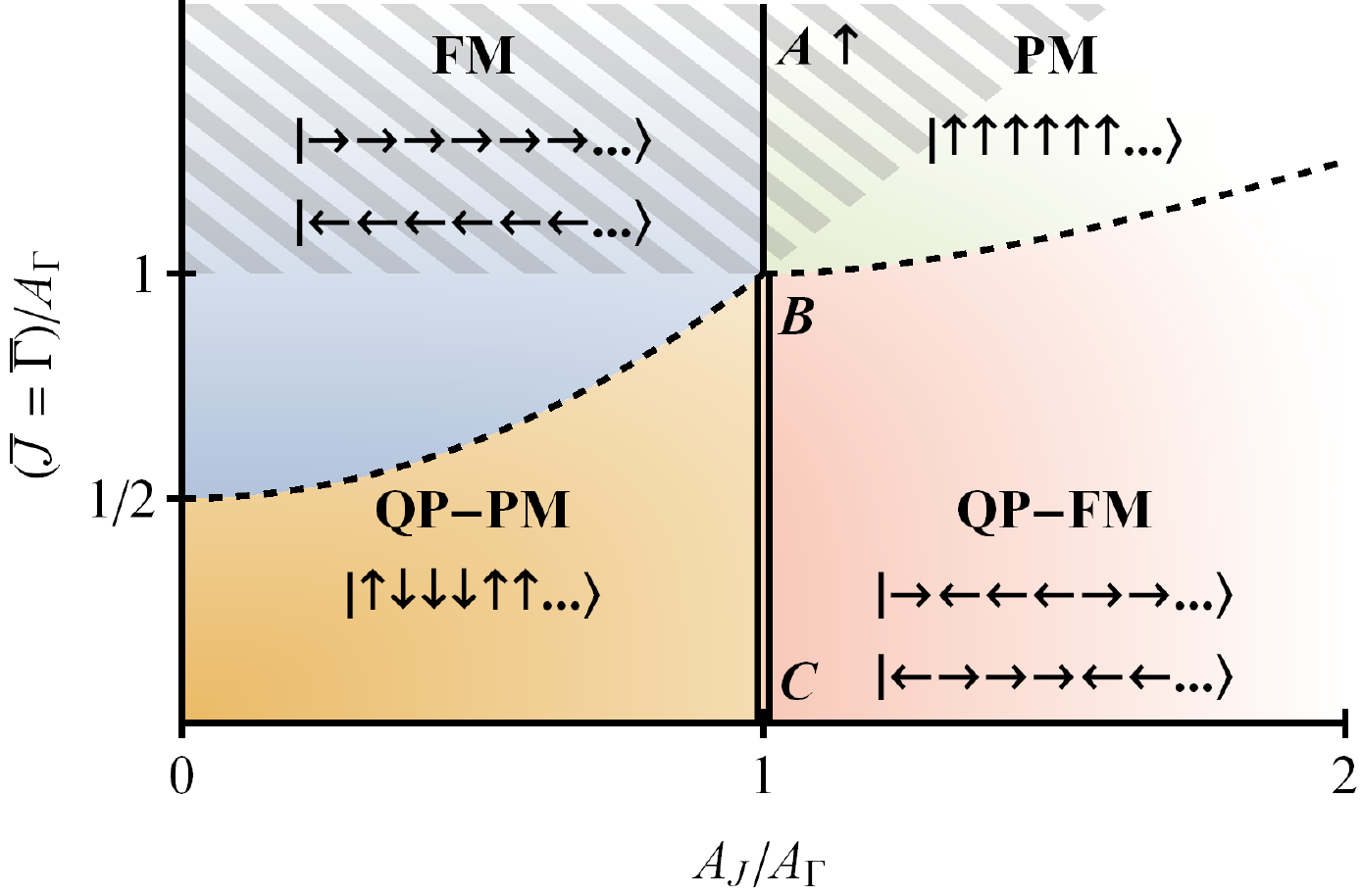}
\caption{
\emph{Phase diagram}. 
The hatched region defines the weakly modulated regime with no weak coupling ($J(Qi), \Gamma(Qi) > 0$ $\forall i$).
The usual gapped ferromagnetic (blue) and paramagnetic (green) phases appear in this regime, separated by a continuous transition in the clean Ising class (segment $AB$).
At stronger modulation, we find two new modulated gapless phases: the QP-PM (yellow), and the QP-FM (red).
The continuous transitions out of these phases (double and dashed lines) are in the new QP Ising class.
}
\label{Fig:FigPhaseD}
\end{center}
\end{figure}

The discovery and growth of quasicrystals~\cite{shechtman1984metallic,levine1984quasicrystals,merlin1985quasiperiodic} motivated the study of critical systems modulated by discrete quasi-periodic substitution sequences~\cite{satija1988quasiperiodic,you1991quantum,vidal1999correlated,hermisson2000aperiodic,hida2001quasiperiodic,tong2002quantum,hida2005renormalization,vieira2005low,vieira2005aperiodic} including the quantum Ising chain~\cite{doria1988quasiperiodicity,igloi1988quantum,tracy1988universality,ceccatto1989quasiperiodic,kolavr1989attractors,benza1989quantum,benza1990phase,lin1992phase,Luck:1993ad,turban1994surface,grimm1996aperiodic,hermisson1997aperiodic,igloi1997exact,igloi1998random,yessen2014properties}. However, recent optical lattice experiments naturally realise smooth quasi-periodic modulation~\cite{roati2008anderson,deissler2010delocalization,schreiber2015observation,bordia2016coupling,luschen2017signatures}. Whilst such modulation has been investigated in related models~\cite{Aubry:1980aa,Azbel:1979aa,Hofstadter:1976aa,Harper:1955kl,thouless1983bandwidths,kohmoto1983metal,kohmoto1987critical,janssen1991characterisation,thouless1991total,hiramoto1992electronic,han1994critical,biddle2009localization,ganeshan2015nearest,gopalakrishnan2017self,iyer2013many,mastropietro2015localization,tezuka2012reentrant,degottardi2013majorana,satija2017topology,devakul2017anderson}, Luck's analysis~\cite{Luck:1993ad} of wandering showed smooth QP modulation to be perturbatively irrelevant at the quantum Ising transition. This deterred further study of the strongly modulated regime, whose physics we here uncover.

The generic QP transverse field Ising model (TFIM) in one dimension has the Hamiltonian:
\begin{align}
\label{Eq:ModelHam}
H &= -\frac{1}{2} \sum_j \J(Qj) \sigma_j^x \sigma_{j+1}^x + \h(Qi) \sigma_j^z, 
\end{align}
where $\J(\theta)$ and $\h(\theta)$ are smooth $2\pi$-periodic functions.
The modulation is quasi-periodic if the wavelength $2\pi/Q$ is an irrational multiple of the lattice length $a=1$.
Our general results apply to a large class of irrational wavevectors (see App.~\ref{App:LogWander}); numerical results use the Golden mean, $Q/2\pi=\tau\equiv (1+\sqrt{5})/2$.
The QP model is best understood as the limit of a sequence of commensurate models with wavevectors $\tilde{Q} = 2\pi p /q$, for coprime integers $p,q$ such that $p/q \to Q/2\pi$ \footnote{Optimal values of $p,q$ are given by the continued fraction approximants to $Q/2\pi$.}.
The period $q$ then acts like a finite length scale which controls scaling behavior.

Using the Jordan Wigner transformation, Eq.~\eqref{Eq:ModelHam} maps on to a free Majorana chain~\cite{lieb1961two}:
\begin{align}
	\label{eq:ham_majorana}
	H &= \frac{i}{2} \sum_j \J(Qj) \gamma_{2j+1} \gamma_{2j +2} + \h(Qj) \gamma_{2j} \gamma_{2j+1}
\end{align}
where $\gamma_i$ are Majorana fermion operators (for conventions and details, see Ref.~\cite{Chandran:2017ab}).
For an open chain in the Ising ordered phase, there are exponentially localised zero modes bound to the system edges. 
The zero mode wavefunction at the left edge is,
\begin{align}
	\label{eq:zeromode_wf}
	\psi^{0}_{2j} \propto \prod_{i<j} \left| \frac{\h(Qi)}{\J(Qi)} \right| \equiv \exp\left(\sum_{i<j} \delta(Qi)\right)
\end{align}
where $\delta(Qi) = \log \left|\h(Qi)/\J(Qi)\right|$ is the local reduced coupling. 
The equation $[\delta(Qi)] = 0$ determines the phase boundary, where $[\cdot]$ denotes spatial averaging.
For QP modulation, the phase boundaries are independent of $Q$ as the spatial average $[\cdot]$ reduces to a phase average $[\cdot]_\theta$.

The couplings in the simplest QP TFIM arise from a single tone: 
\begin{align}
\J(Qi) & = \BJ + \AJ \cos(Q(i+1/2) + \phi) \nonumber\\
\h(Qi) & = \Bh + \Ah \cos(Qi+\phi+\Delta)
\label{eq:SineMod}
\end{align}
where the phases $\phi$ and $\Delta$ shift the couplings with respect to the lattice.
We highlight an interesting slice of the phase diagram in Fig.~\ref{Fig:FigPhaseD} where $\BJ = \Bh$.
There are four phases.
The usual gapped Ising PM and FM phases arise in the weakly modulated regime ($\BJ = \Bh > \AJ,\Ah$) at the top of the figure.
Two new phases appear at strong modulation, when the couplings take both positive and negative signs:  
a QP-FM with modulated ferro- and anti-ferromagnetic correlations, and a QP-PM with modulated transverse magnetization.

The two QP phases are gapless with localised excitations at all energies. 
Heuristically, this is a consequence of weak couplings (of order $1/q$) which occur when $Qi$ in Eq.~\eqref{eq:SineMod} samples near the zeros of $\J(\theta)$ or $\h(\theta)$.
The weak couplings nearly cut the chain which localises excitations on either side.
In turn, excitations localised on the weak links have arbitrarily low energy as $q\to \infty$. 
We note that the gapless excitations are not associated with rare regions, unlike in the Griffiths-McCoy phase of the disordered Ising chain \cite{griffiths1969nonanalytic,McCoy:1968aa,mccoy1969theory}.

\begin{figure}
\centering
\includegraphics[width=\columnwidth]{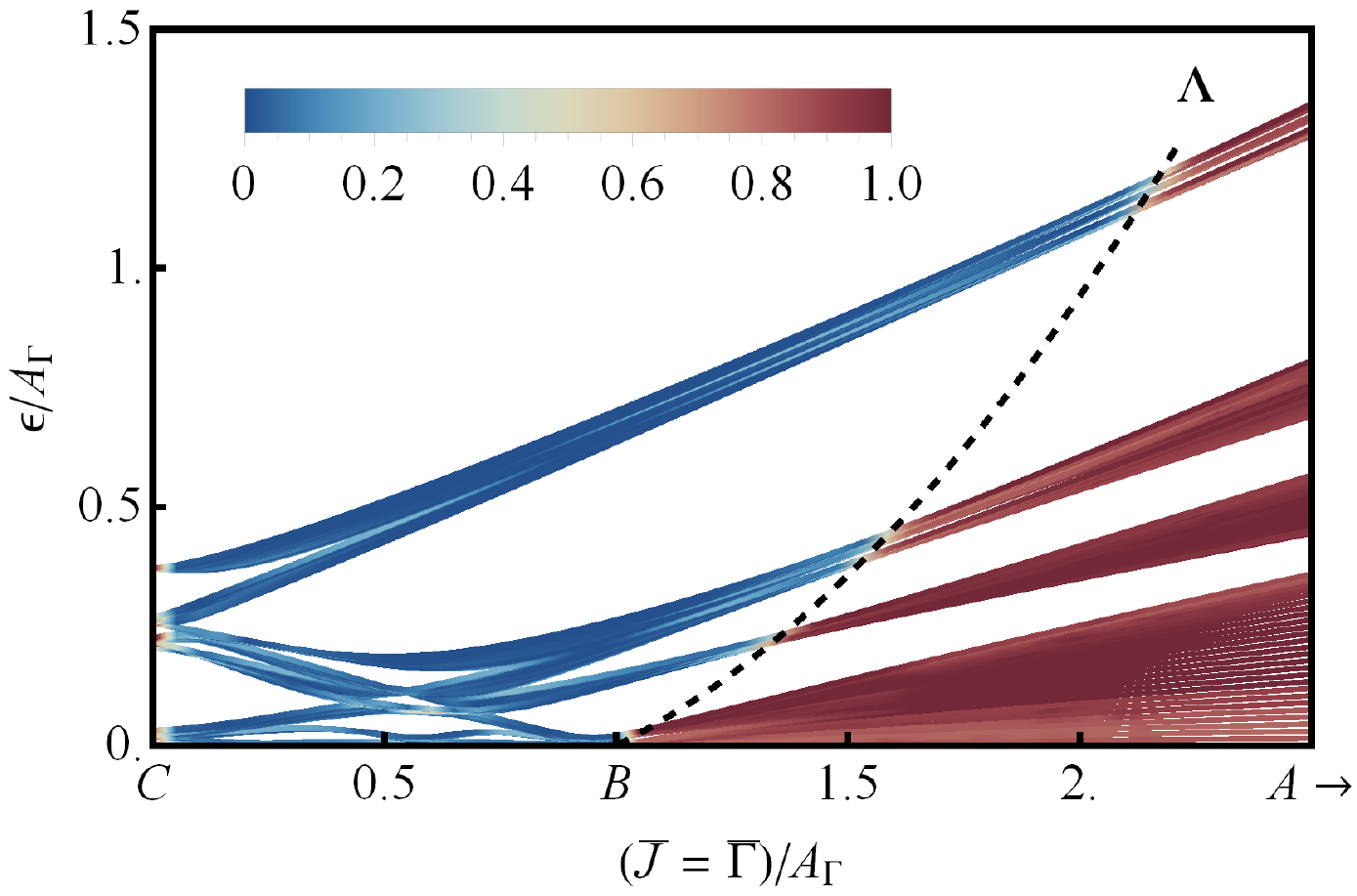}
\caption{
\emph{Localisation properties of excitation spectrum on line $ABC$}. 
The low-energy excitations on the segment $AB$ are extended (red) up to a finite energy cut-off $\Lambda$, above which they are localised (blue).
The cutoff $\Lambda$ vanishes at the multi-critical point $B$ so that all finite energy excitations are localised on the segment $BC$. 
The color quantifies the scaling of the inverse participation ratio $I = \sum_i |\psi_i|^4 \sim q^{-a}$; $a=0$ ($1$) for localised (extended) states. Parameters: $q = 233$, $\Delta = 42 \pi/233$, $\phi = \sqrt{2}$
}
\label{fig:mobedge}
\end{figure}

In what follows we focus on the phase boundary $\AJ = \Ah$ (segment $ABC$, Fig.~\ref{Fig:FigPhaseD}).
All of the points on this line are Ising self-dual and accordingly critical. 
QP modulation is perturbatively irrelevant at the clean Ising transition \cite{Luck:1993ad}.
Our numerics (not shown) confirm that all critical exponents in the weak modulation regime (segment $AB$, Fig.~\ref{Fig:FigPhaseD}) are consistent with clean universality.
The difference between the unmodulated model and the weakly modulated model become apparent only at high energy:
Fig.~\ref{fig:mobedge} shows that the low energy excitations are extended (red) up to a finite cutoff energy $\Lambda$,  above which they become localised (blue).
This mobility edge collapses ($\Lambda \to 0$) at the multicritical point $B$. 
On the segment $BC$, all finite energy excitations are localised, consistent with the localisation of the adjacent QP-PM and QP-FM phases. 
This is our first qualitative indication that the critical properties of the QP and clean transitions are quite different.

Before turning to the detailed properties of the QP Ising transition, we briefly review single parameter scaling. 
At clean critical points, coarse-grained observables are scale free \cite{Goldenfeld:1992aa}. 
Single parameter scaling posits that a single length scale and corresponding time scale diverge at the transition, 
\begin{align}
	\xi \sim [\delta]^{-\nu}, \qquad \xi_t \sim \xi^z
    \label{eq:xi}
\end{align}
where $\nu$ and $z$ are the correlation length and dynamical critical exponents, respectively.
These control the long distance and long time correlations in the vicinity of the critical point. 
For example,
\begin{align}
	\label{eq:xx_scalingform}
	[\langle \sigma_i^x(t) \sigma_{i+r}^x(0)  \rangle] \sim \frac{1}{|r|^{2 \Delta_{\sigma}}} \mathcal{C}(r /\xi, t / \xi_t)
\end{align}
where $\Delta_\sigma$ is a scaling dimension and $\mathcal{C}$ a scaling function.
These are both part of the universal data of the critical point.
It is well known that the scaling ansatz holds at the clean Ising transition.

In the disordered and QP transitions, the scaling ansatz needs to be refined. 
The \emph{spatially averaged} correlation functions satisfy scaling in the form of Eq.~\eqref{eq:xx_scalingform} with a single \emph{mean} correlation length $\xi$. 
However, the \emph{typical} correlation functions may decay on a shorter, but still divergent, length scale $\xi_{\typ} \sim [\delta]^{-\nu_{\typ}} \ll \xi$.
Fisher~\cite{fisher1995critical} first emphasized this in the disordered case, where $\nu_{\typ}=1$ while $\nu = 2$. 
In the QP case, we will find a much weaker logarithmic separation between $\xi_{\typ}$ and $\xi$.

An additional wrinkle for the disordered and QP Ising transitions is that they separate phases in which all excitations are localised. 
The localisation length $\xi_{\mathrm{loc}}$ is a function of energy $\epsilon$ and deviation $[\delta]$ which must diverge as $[\delta],\epsilon \to 0$. 
By scaling, we can compute $z$ from the energy dependence of $\xi_{\mathrm{loc}}$ and $\nu$ from its $[\delta]$ dependence.
Henceforth, we drop the subscript from $\xi_{\mathrm{loc}}$ as it coincides with $\xi$ where they are both defined.

\begin{figure*}
\centering
\includegraphics[width=0.33\textwidth]{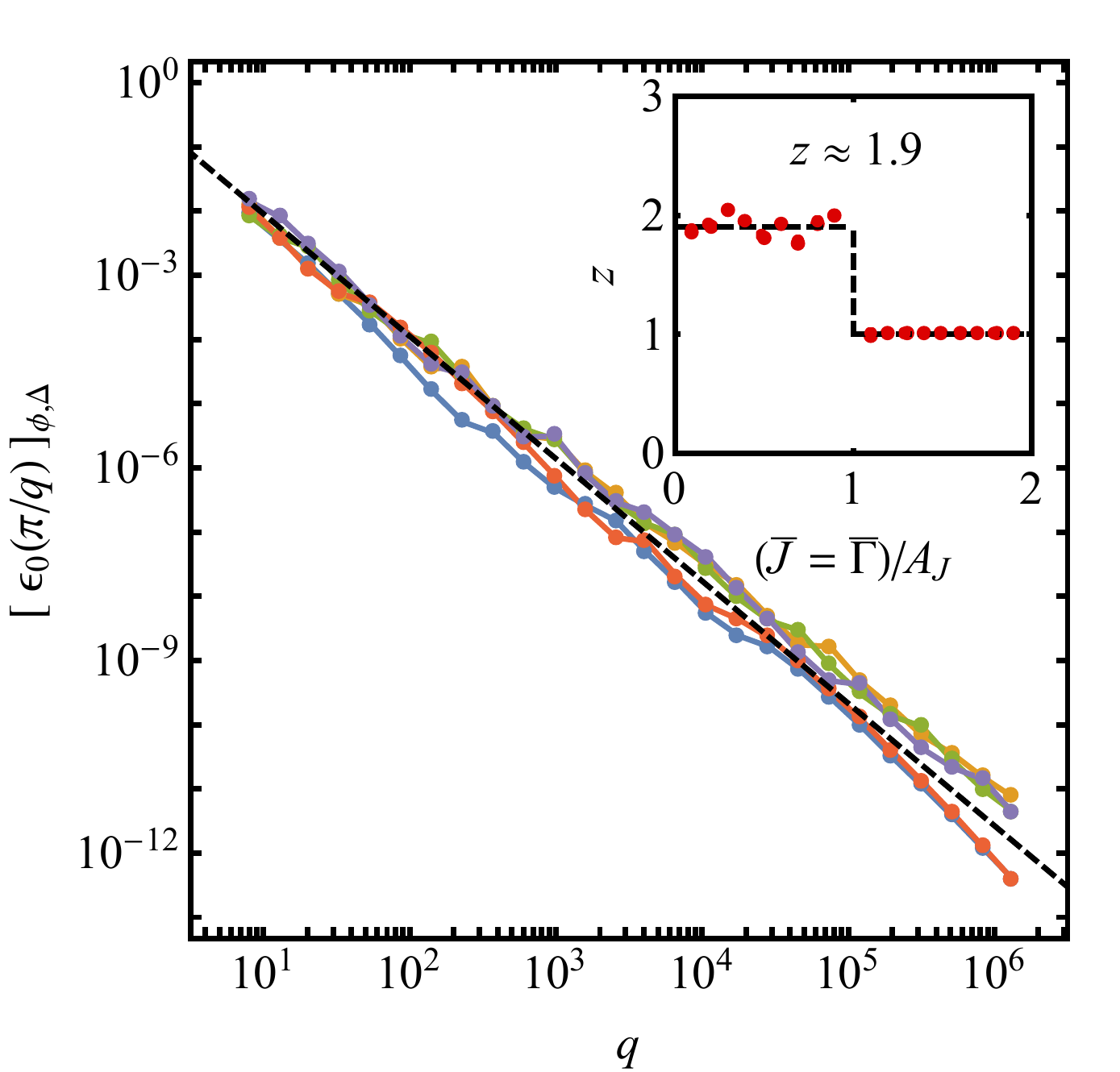}
\includegraphics[width=0.326\textwidth]{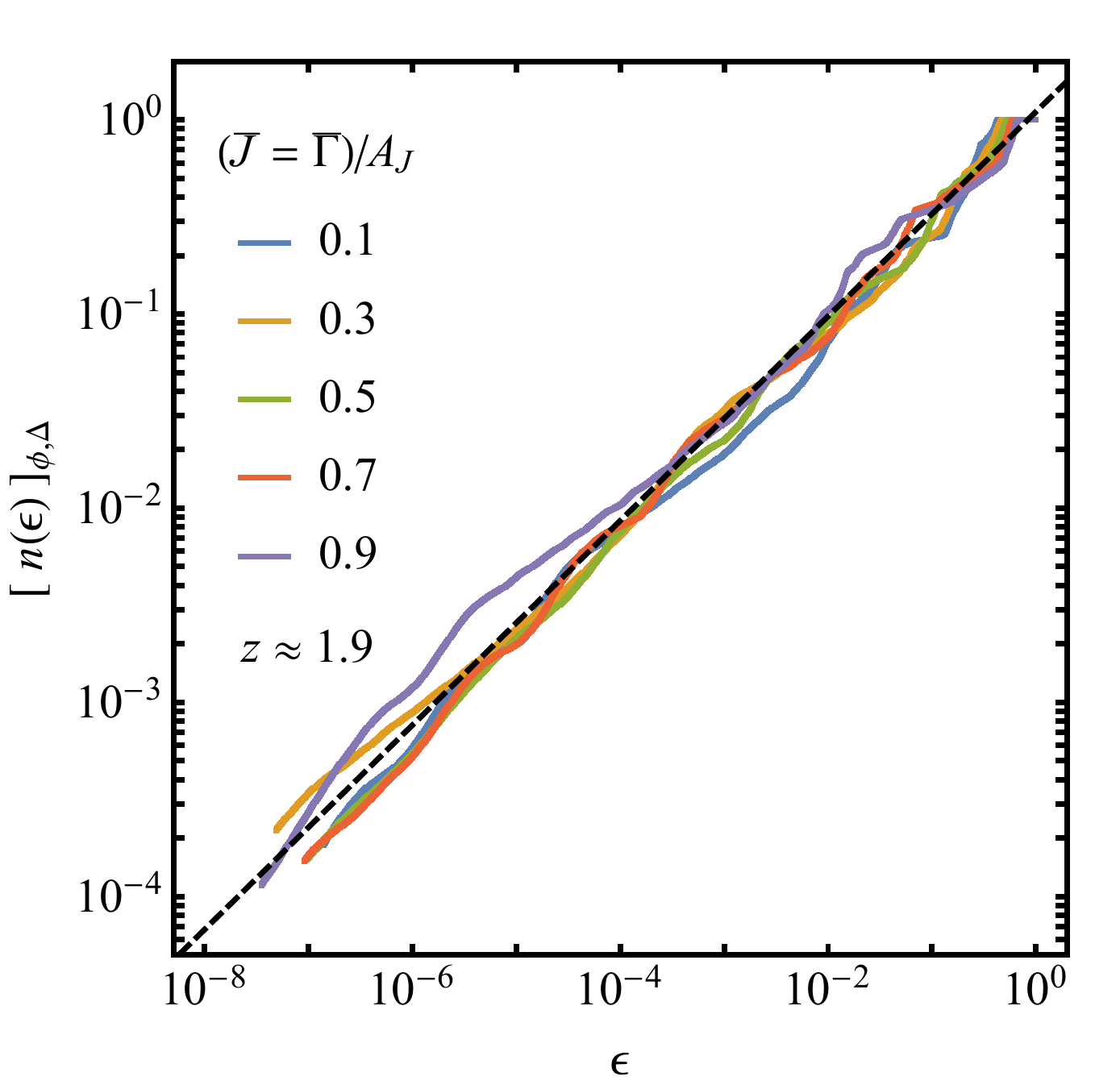}
\includegraphics[width=0.328\textwidth]{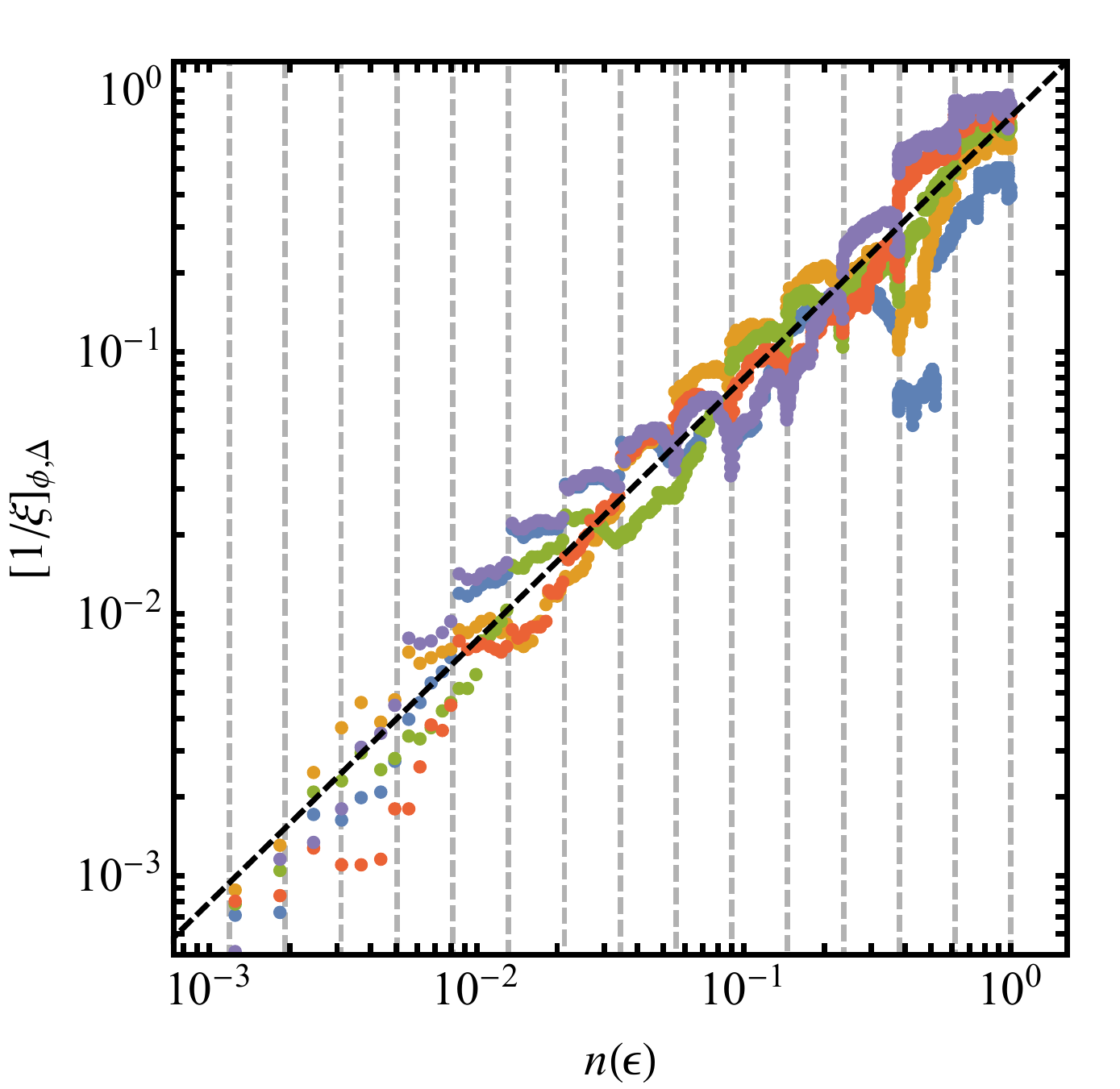}

\caption{
\emph{Dynamical scaling at the QP Ising transition}:
(left) The maximum energy of the lowest band $\cexp{\epsilon_0}_{\phi,\Delta}$ scales as a power law $q^{-z}$ over 5 orders of magnitude (mean over $5000$ $\phi,\Delta$ samples at each Fibonacci length).
(inset) Least squares fit exponent $z$ as a function of parameter along the phase boundary.
(centre) The integrated density of states $[n(\epsilon)]_{\phi,\Delta} \sim \epsilon^{1/z}$ over 6 orders of magnitude in energy at the largest size available ($q=4181$).
(right) The inverse localisation length $[1/\xi(\epsilon)]_{\phi,\Delta}$ is linearly proportional to $n(\epsilon)$, consistent with single parameter scaling. 
The deviations from the central trend show sharp features at the log-periodically spaced convergents of the golden ratio $\tau$ (vertical dashed lines). 
In all panels, the measurements are shown at 5 different values of $\Bh/\Ah$ on segment $BC$ of Fig.~\ref{Fig:FigPhaseD}, indicating universality.
Standard errors are smaller than point size; deviations from power law trends are deterministic and due to the QP modulation. 
}
\label{fig:Z}
\end{figure*}

We now turn to analytic and numerical computation of the critical properties of the QP Ising transition. 

\emph{Typical correlations---} 
We begin with the exponent that controls the decay of the zero energy wavefunction (Eq.~\eqref{eq:zeromode_wf}) across a region of size $\ell$:
\begin{align}
	S_\ell(i) \equiv \log \left| \frac{\psi^0_{2(i+\ell)}}{\psi^0_{2i}} \right| = \sum_{j=i}^{i+\ell-1} \delta(Qj)
\label{eq:SL}
\end{align}
The typical correlation length follows immediately from evaluating the typical exponent controlling decay: $[S_\ell] = \ell [\delta] \sim \ell / \xi_{\typ}$.
From Eq.~\eqref{eq:xi} this implies  $\nu_{\typ}=1$. 

\emph{Mean correlations---}
The spatial modulation induces fluctuation in the exponent $S_\ell(i)$, which are characterized by the scale dependent variance (`wandering'),
\begin{align}
\label{eq:variance_def}
	\sigma^2(S_\ell) = [S_\ell^2] - [S_\ell]^2.
\end{align}
If the wandering $\sigma > |[S_\ell]|$, then the system has a density of regions of size $\ell$ in which it is locally in the opposite phase. 
Thus, the spatially averaged correlations at this scale cannot determine the global phase; this generalizes the Harris-Luck instability argument \cite{Harris:1974aa,Luck:1993fu} to the strong modulation regime.
Furthermore $\sigma(S_{\xi}) \sim |[S_{\xi}]|$ defines the \emph{mean} correlation length $\xi$ above which the global phase is determined.
As $[\delta] \to 0$, $\xi$ diverges faster than $\xi_{\typ}$ if the wandering grows with $l$. 

For disordered chains, the exponent $S_\ell$ undergoes a random walk so that $\sigma \sim \sqrt{\ell}$.
In the QP chain, the long-range correlations of the spatial modulation produce a more complicated non-monotonic wandering (see Fig.~\ref{fig:bball}).
In particular, there are exponentially separated special lengths $\ell$ (the convergents of $Q/2\pi$) at which $\sigma$ is anomalously small.
Nevertheless, for typical large $\ell$, the wandering $\sigma^2$ is very close to its Cesaro average:
\begin{align}
\label{eq:dub_w}
	\frac{1}{\ell}\sum_{\ell'=1}^\ell \sigma^2(S_{\ell'}) \sim \left\{ \begin{array}{lll} 
	c && \mathrm{if\ } |J(\theta)|,|\Gamma(\theta)| > 0 \\
	w \log \ell && \mathrm{otherwise} 
	\end{array} \right.
\end{align}
The two cases in Eq.~\eqref{eq:dub_w} are physically distinguished by the presence of weak couplings, and correspond to segments $AB$ and $BC$ in Fig.~\ref{Fig:FigPhaseD}, respectively.
Here, $c$ is an $l$-independent constant and we pithily dub $w$ the \emph{logarithmic wandering coefficient} (see Appendices~\ref{App:LogWander},~\ref{App:wGolden} for derivation). 
Generically, this coefficient only depends on the wavevector $Q$ and number and order of the zeros within a period of the coupling functions. 
We conjecture that $w$ uniquely parametrizes the family of QP Ising transitions.

The correlation length exponent follows immediately from the coarse grained wandering described by Eq.~\eqref{eq:dub_w}. 
On segment $AB$, $\nu = 1$ and the mean and typical correlations do not separate. 
This is consistent with $AB$ being in the clean Ising universality class~\cite{Luck:1993ad}. 
On segment $BC$, the mean correlation length is logarithmically enhanced,
\begin{align}
\label{eq:ximean_log}
\xi \sim [\delta]^{-1} \log^{1/2}(1/[\delta])
\end{align}
compared to $\xi_{\typ}$ (i.e., ``$\nu = 1^+$'').

\emph{Dynamical exponent---}
The dynamic properties show more dramatic signatures of the change in universality. 
Treating the secular equation of the Hamiltonian~\eqref{eq:ham_majorana} to leading order in the wandering of $S_\ell$, we find,
\begin{align}
z \approx 1+w
\end{align}
This follows from estimating the scaling of the bandwidth of the lowest band with period $q$ (see App.~\ref{App:zEstimate}) \footnote{We measure all energies relative to $\epsilon = 0$}. 
For the Golden mean, $Q/2\pi = \tau$, the wandering coefficient $w = \frac{2\pi^2}{15\sqrt{5} \log \tau} \approx 1.2$ \cite{speyerprivate}, which produces an estimate $z \approx 2.2$.

This estimate of $z$ neglects spatial correlations of the wandering, higher order moments and the deterministic deviations of $\sigma(S_\ell)$ from its Cesaro average. 
We are thus unable to detect multiplicative logarithmic corrections to the dynamical scaling which are suggested by Eq.~\eqref{eq:ximean_log}. 
All results which follow are only valid up to the possibility of such corrections. 

Figure~\ref{fig:Z} shows three different numerical measurements of $z$ which collectively verify both single parameter scaling and universality. 
Panels (a) and (b) probe dynamical scaling through the $\phi, \Delta$ averaged integrated density of states $n(\epsilon) \sim \epsilon^{1/z}$ at asymptotically vanishing and finite energy scales, respectively.
With periodic modulation $q$, the maximum energy $\epsilon_0$ of the lowest miniband satisfies $n(\epsilon_0) = 1/q$. 
This implies $\epsilon_0 \sim q^{-z}$; panel (a) confirms this power law holds with exponent $z\approx 1.9$ for system sizes over 5 orders of magnitude up to $q \approx 10^6$.
Panel (b) shows that the same exponent governs the scaling of $n(\epsilon)$ with $\epsilon$ up to finite energy.
Here, $n(\epsilon)$ is extracted from the histogram of energy levels from $10^4$ diagonalisations at size $q = 4181$ across sampled values of $\phi, \Delta$. 
Both panels collapse data from a series of points along the $BC$ phase boundary, consistent with universality.

We extract $\xi^{-1}(\epsilon)$ from a least squares fit to the relationship
\begin{equation}
\log \left[ \left|  \psi_i(\epsilon) \bar\psi_{i+r}(\epsilon)\right| \right] = - r \xi^{-1}(\epsilon)  + \textrm{const}
\end{equation}
where $\psi_i(\epsilon)$ is the eigenmode at energy $\epsilon$ for systems of size $q=1597$.
We again see evidence of universality along the phase boundary.

In all three panels of Fig.~\ref{fig:Z} the visible deviation from pure power laws reflect deterministic modulation.
The phase averaging of various quantities reduces the deviations from the central trends but does not completely suppress them.
We expect deviations from pure power laws due to rare values of $l$ at which $\sigma(S_\ell)$ deviates significantly from its Cesaro mean (see Eq.~\eqref{eq:dub_w}). 
These special values are marked by dashed lines in panel (c) where they correlate with atypically delocalised excitations. 

The presence of these special points leads us to conjecture that the single parameter scaling forms, eg. in Eq.~\eqref{eq:xx_scalingform}, hold up to a non-universal multiplicative $O(1)$ function. 
That is, the scaling form provides the \emph{envelope} for these $O(1)$ fluctuations. 
A consequence of this hypothesis is that the critical exponents are well-defined as $q \to \infty$ but the convergence of finite-size numerical estimates is only $O(1/\log q)$. 
This is consistent with the scatter in the inset of panel (a) in Fig.~\ref{fig:Z}.

\begin{figure}
\centering
\includegraphics[width=0.47\textwidth]{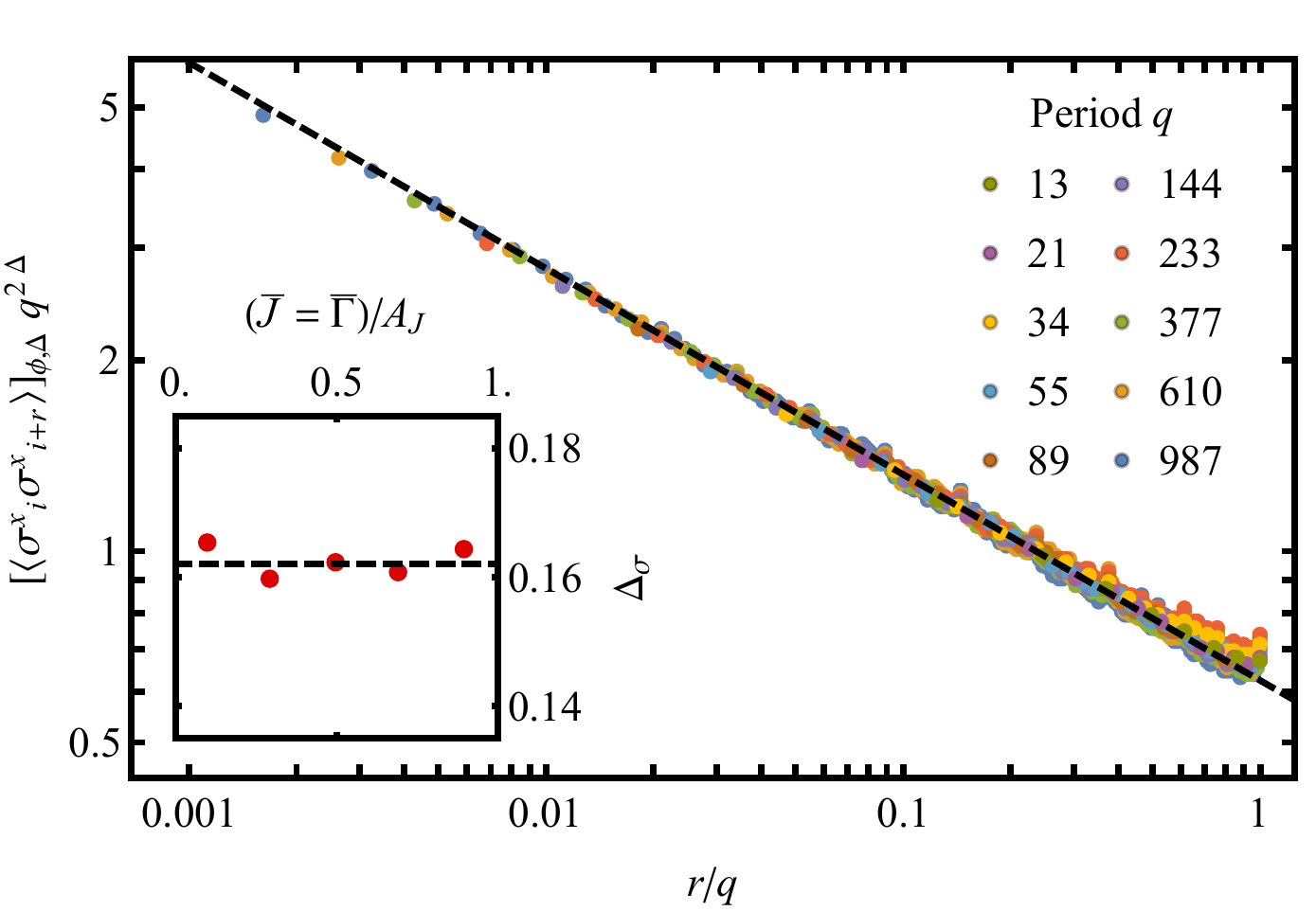}
\caption{
\emph{Finite size scaling of spin correlations:} 
(main) The average spin correlations $q^{2 \Delta_\sigma}[\left<\sigma_i^x\sigma_{i+r}^x\right>]_{i, \phi, \Delta} $ collapse when plotted versus fractional separation $r/q$ for critical dimension $\Delta_\sigma \approx 0.16$ at $(\J = \h)/\AJ = 0.5$.  
(inset) Least median deviation fit exponent $\Delta_\sigma$ is stable along the segment $BC$ consistent with universality.
}
\label{fig:spinspin}
\end{figure}

\emph{Scaling dimensions---}
The equal time correlators at the QP Ising transition decay with a faster power law than at the clean Ising transition, but slower than at infinite randomness.
Fig.~\ref{fig:spinspin} shows the excellent finite-size scaling collapse of the mean equal time spin correlator $[\langle \sigma^x_i \sigma^x_{i+r} \rangle]_{i,\phi, \Delta}$ at the QP transition. Using data from different points on the QP transition line we extract $\Delta_\sigma \approx 0.16$ (see Eq.~\eqref{eq:xx_scalingform}).
We find similarly enhanced value of the scaling dimension of the Majorana fermions $\Delta_{\gamma} \approx 0.63$ (data not shown).
In contrast, for the clean TFIM $\Delta_\sigma = 0.125$, $\Delta_\gamma = 0.5$, and for the random TFIM $\Delta_\sigma = (3-\sqrt{5})/4 \approx 0.19$, $\Delta_\gamma \approx 1.1$~\cite{fisher1995critical,igloi2017transverse}.

The QP critical correlations are observed on length scales $r < q$; for $r > q$, the system is actually periodic and we recover clean Ising correlations~\footnote{Transition to Ising criticality for $r > q$ occurs only if the system is clean Ising critical. Due to $\mathrm{O}(1/q)$ finite-size corrections to the clean Ising critical boundaries this is not automatically satisfied. For $q\Delta - p \pi \in 2 \pi \mathbb{Z}$ the QP Ising boundary is a clean Ising boundaries. These $\Delta$ form a dense set in  $[0,2 \pi]$ in the QP limit.}.
In Fig.~\ref{fig:spinspin}, this is presaged by the small upturn near $r=q$.

\emph{Discussion---}
Weak quasi-periodic modulation is well-known to be perturbatively irrelevant at the clean Ising transition~\cite{Luck:1993ad}. 
We have shown that sufficiently strong modulation destabilizes this transition and drives the TFIM into a new spatially modulated QP Ising transition. 
Like in the infinite randomness case, the low energy excitations are localised throughout the critical fan, although with a power law diverging localisation length as $\epsilon \to 0$.
The exponents of the QP Ising transition lie between their clean and disordered counterparts.
The most dramatic signatures of this transition are in the localised dynamics and larger specific heat as compared to the clean case.

Our results rely on the emergence of logarithmic wandering with coefficient $w$ describing the dominant long-distance fluctuations of the order.
We conjecture that $w$ controls the universal content of a family of QP Ising transitions. 
As $w$ is only a function of wavenumber $Q$ and the number and order of the zeros of $J(\theta), \Gamma(\theta)$, it follows that the critical properties are insensitive to smooth perturbations which preserve the wavenumber. 
We have provided numerical evidence for this universality by varying couplings along the boundary $BC$.

Remarkably, logarithmic wandering arises \emph{without weak couplings} when $\J(\theta), \h(\theta)$ have step discontinuities.
Technically, this follows from the $1/k$ tails in the Fourier transform of $\delta(\theta)$.
As the size of the steps controls $w$, we can realize a large family of QP Ising transitions with tunable exponents in such models.
Some aspects of this have been previously discussed in connection with substitution sequence modulation in the TFIM~\cite{doria1988quasiperiodicity,igloi1988quantum,tracy1988universality,ceccatto1989quasiperiodic,kolavr1989attractors,benza1989quantum,benza1990phase,lin1992phase,Luck:1993ad,turban1994surface,grimm1996aperiodic,hermisson1997aperiodic,igloi1997exact,igloi1998random,yessen2014properties}.

The stability of the QP Ising transitions to the introduction of interactions is an open question.
On the one hand, interactions which effectively lift weak couplings could destroy the log wandering.
On the other hand, the example of step modulation suggests that weak couplings are not strictly necessary for modified criticality.

\begin{acknowledgments}

We are grateful to David Speyer for useful correspondence and for the final calculation of App.~\ref{App:wGolden}, we are further grateful to B. Altshuler, Y.Z. Chou, M. Foster, D. Huse, B. McCoy, J.H. Pixley, and S. Sondhi for useful discussions. We are grateful the Simons Center for Geometry and Physics (SCGP) and the Kavli Institute for Theoretical Physics (KITP) in Santa Barbara (supported by National Science Foundation grant No. NSF PHY11-25915), for their hospitality, and to the Shared Computing Cluster (administered by Boston University Research Computing Services) which was used for computational work. C.R.L. acknowledges support from the Sloan Foundation through a Sloan Research Fellowship and the NSF through grant No. PHY-1656234. Any opinion, findings, and conclusions or recommendations expressed in this material are those of the authors and do not necessarily reflect the views of the NSF.
\end{acknowledgments}

\appendix

\section{Logarithmic wandering when $Q$ is a badly approximable number}
\label{App:LogWander}

\begin{figure}
\centering
\includegraphics[width=0.47\textwidth]{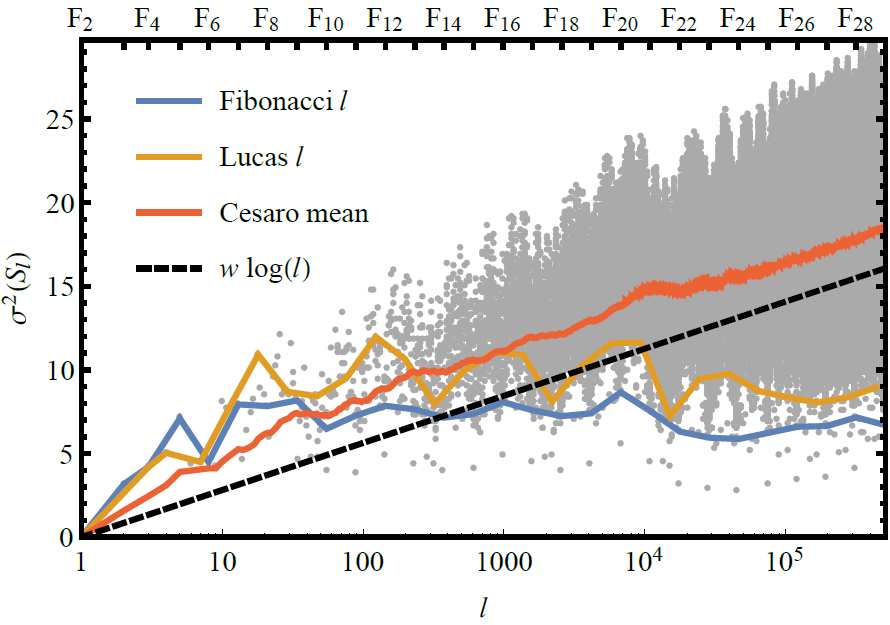}
\caption{
\emph{Logarithmic wandering $\sigma^2(S_\ell)$:} The infimum and supremum of $\sigma^2(S_\ell)$ (grey dots) separate in the limit of large $\ell$. The supremum, and generic values, scale as $\log \ell$ while the infimum is a constant. The Ces\'{a}ro mean~\eqref{eq:AppCL} (red), converges to $w \log \ell$ (black, dashed). Two series for which $\sigma^2(S_\ell)$ is bounded by a constant, the Fibonacci numbers (blue), and Lucas numbers (gold), are highlighted. Fibonacci numbers $F_n$ are marked on the upper horizontal axis. Parameters $\J = \h = 1$, $\AJ = \Ah = 2$, $\phi=\sqrt{2}$, $\Delta=2.261$, and $Q/2\pi = \tau$ the Golden ratio.}
\label{fig:bball}
\end{figure}

In this appendix we study the asymptotic behaviour of
\begin{equation}
\sigma^2(S_\ell) = \cexp{S_\ell^2(i)} - \cexp{S_\ell(i)}^2
\label{eq:AppSL}
\end{equation}
where $S_\ell(i)$ is defined as in~\eqref{eq:SL} and $\cexp{\cdot}$ denotes the spatial average. We show that (in the sense of the Ces\'{a}ro mean) this quantity grows logarithmically in $\ell$ when $Q/2\pi$ is a badly approximable number (defined below).

Throughout these appendices we use the usual notations of Bachmann-Landau and Knuth to denote asymptotic relations: in the limit of large $\ell$, $f_\ell \sim g_\ell$ indicates
\begin{equation}
\lim_{\ell \to \infty} \frac{f_\ell}{g_\ell} = 1,
\end{equation}
whilst both $f_\ell = O(g_\ell)$  and $g_\ell = \Omega(f_\ell)$ indicate that $|f_\ell|< c |g_\ell|$ for some constant $0<c<\infty$ at sufficiently large $\ell$. Finally $f_\ell = \Theta(g_\ell)$ implies both $f_\ell = O(g_\ell)$ and $f_\ell = \Omega(g_\ell)$.

As seen in Fig.~\ref{fig:bball} the $O(\ell^0)$ infimum and the $O(\log \ell)$ supremum of the sequence $\sigma^2(S_\ell)$ (gray) are asymptotically separated. Small $O(\ell^0)$ values of $\sigma^2(S_\ell)$ are found for $\ell$ in certain exponentially spaced sub-sequences, such as the Fibonacci numbers $F_n$ and the Lucas numbers $L_n$. As $\ell \to \infty$ there are infinitely many such sparse sub-sequences in which $\sigma^2(S_\ell)$ is bounded by a finite constant. For these reasons we characterise the typical behaviour of the sequence by the Ces\'{a}ro mean \begin{equation}
\cexp{\sigma^2(S_\ell)}_\textrm{Ces\'{a}ro} = \frac{1}{\ell} \sum_{\ell'=1}^{\ell} \sigma^2(S_{\ell'}).
\label{eq:AppCL}
\end{equation}
For strong smooth modulation (i.e. for smooth functions $\J(\theta)$, $\h(\theta)$ with at least one zero) this quantity scales as 
\begin{equation}
\cexp{\sigma^2(S_\ell)}_\textrm{Ces\'{a}ro} \sim w \log \ell
\label{eq:AppWander}
\end{equation}
for for the logarithmic wandering coefficient $w$ while for weak smooth modulation (i.e. smooth $\J(\theta)$, $\h(\theta)$ bounded away from zero) $\sigma^2(S_\ell)$ is bounded by a constant.

The logarithmic scaling~\eqref{eq:AppWander} holds for strong modulation when $\zeta=Q/2\pi$ is in a certain subset of irrational numbers, the ``badly approximable numbers''. These irrationals, when written as a continued fraction
\begin{equation}
\zeta = a_0 + \frac{1}{a_1 + \frac{1}{a_2+ \ldots}},
\end{equation}
have partial quotients $a_n$ which are finitely bounded: $a_n \leq a_\mathrm{max} < \infty$. Equivalently, $\zeta$ is badly approximable if and only if there exists a finite constant $x_0$ such that
\begin{equation}
|k \fracp{k \zeta}| \geq x_0 >0
\label{eq:BadlyApprox}
\end{equation}
for all integers $k$~\cite{hindry2013diophantine}. Here $\fracp{\alpha} = \alpha - \textrm{Round}(\alpha)$ denotes the fractional part of the real number $\alpha$ and $\textrm{Round}(\alpha)$ is the nearest integer to $\alpha$. As an example, the quadratic integers are a subset of badly approximable numbers.

In App.~\ref{App:wGolden} evaluate $w$ exactly for the case of the quadratic integer $Q/2\pi = \tau = (1+\sqrt{5})/2$, the Golden ratio. In the rest of this appendix we show that this scaling holds for badly approximable numbers in general.

To show~\eqref{eq:AppWander} it is most natural to work in Fourier space. The Fourier representation of the reduced couplings is given by 
\begin{equation}
\delta(\theta) = \sum_k \hat\delta_k \mathrm{e}^{\mathrm{i} k \theta}.
\end{equation}
Using the definition of $S_\ell$ in~\eqref{eq:SL} and the equivalence of phase and spatial averages
\begin{equation}
\begin{aligned}
\sigma^2(S_\ell) &= \sum_{k\neq 0} |\hat\delta_k|^2 \frac{\sin^2(\Q k \ell/2)}{\sin^2(\Q k /2)}.
\end{aligned}
\label{eq:SigEll}
\end{equation}
Adopting a shorthand $C_\ell := \cexp{\sigma^2(S_\ell)}_\textrm{Ces\'{a}ro}$ then
\begin{equation}
C_\ell = \sum_{k=1}^{\infty} |\delta_k|^2 f_\ell(Q k/2)
\label{eq:Cl}
\end{equation}
follows from~\eqref{eq:AppCL} and~\eqref{eq:SigEll}, where the kernel
\begin{equation}
f_\ell(\theta) = \frac{1+2\ell}{2 \ell} \frac{1}{\sin^2\theta} \left(1 - \frac{\sin((1+2 \ell)\theta)}{(1+2 \ell)\sin \theta} \right).
\label{eq:fl}
\end{equation}
See Fig.~\ref{fig:fl} for an illustration of $f_\ell$ in its unit cell. For weak modulation $\delta(\theta)$ is smooth and hence $|\delta_k|$ decays exponentially in $k$. In this case~\eqref{eq:Cl} converges to a constant. For strong smooth modulation
$|\delta_k|^2 = O(k^{-2})$ (Eq.~\eqref{eq:AppDelkFinal}) and furthermore, from~\eqref{eq:dkcesaro} $\cexp{|\delta_k|^{-2} k^{2}}_\textrm{Ces\'{a}ro} = 1$. If we neglect pathological cases where small values of $\delta_k$ coincide precisely with large values of $f_\ell(Q k/2)$, then
\begin{equation}
C_\ell \sim \sum_{k=1}^{\infty}\frac{f_\ell(Q k/2)}{k^2}.
\label{eq:CL2}
\end{equation}
The sum of terms for $k>\ell$ can be bounded by a constant (see App.~\ref{App:bound}), so any asymptotic growth must come from the terms $k \leq \ell$
\begin{equation}
C_\ell \sim \tilde{C}_\ell := \sum_{k=1}^{\ell}\frac{f_\ell(\pi \zeta k)}{k^2}.
\label{eq:Ctwiddle}
\end{equation}
where $\zeta = Q/2\pi$. In the remaining parts of this section, we show in App.~\ref{sec:Clesssim} that $\tilde{C}_\ell = O( \log \ell)$ and in App.~\ref{sec:Cgtrsim} we show $\tilde{C_\ell} = \Omega(\log \ell)$. From these it follows that
\begin{equation}
C_\ell = \Theta (\log \ell) 
\label{eq:Ctheta}
\end{equation}
and hence that, for generic values of $\sigma$ (in the sense of the Ces\`{a}ro mean) $\sigma^2(S_\ell)$ grows logarithmically. So far in reaching~\eqref{eq:Ctheta} we have analysed only the scaling, and are unable to comment on whether the upper and lower bounds implied by the $\Theta$ converge. However, as $\sigma^2(S_\ell) = O(\log\ell)$ (see App.~\ref{app:sigmabound}) scales with the same bound, it would require increasingly large sequences of uncharacteristically small or large values for $C_\ell$ to be non convergent. Numerically we find this is not the case (see Fig.~\ref{fig:bball}) and that $C_\ell$ converges
\begin{equation}
C_\ell \sim w \log \ell.
\end{equation}
A final comment is in order regarding an assumption we have made: In going from~\eqref{eq:Cl} to~\eqref{eq:CL2} we have neglected special cases where $\delta_k$ becomes correspondingly small at precisely the values at which values $f_\ell(Q k /2 )$ becomes large: when $Q k$ is close to $2 \pi \mathbb{N}$, (see Fig.~\ref{fig:fl}). In such cases when the denominator in~\eqref{eq:SigEll} is small the numerator is correspondingly small, altering the scaling of the sum. Such fine tuning causes quasi-periodic modulation the logarithmic wandering to be lost, $\sigma^2(S_\ell)$ is bounded by a constant, and the clean Ising universality is realised. This is the case for example if $\delta_k = \sin(Q k/2)/k$. Such a situation is not stable to perturbations that break this fine tuning. An example of a situation with such fine tuning is modulation by the Fibonacci word, which is found to be irrelevant~\cite{igloi1988quantum,ceccatto1989quasiperiodic,benza1989quantum,hermisson1997aperiodic} despite that $|\delta_k|^2 = O(k^{-2})$.

\subsection{Asymptotic upper bound on $\tilde{C}_\ell$:}
\label{sec:Clesssim}

\begin{figure}
\centering
\includegraphics[width=0.47\textwidth]{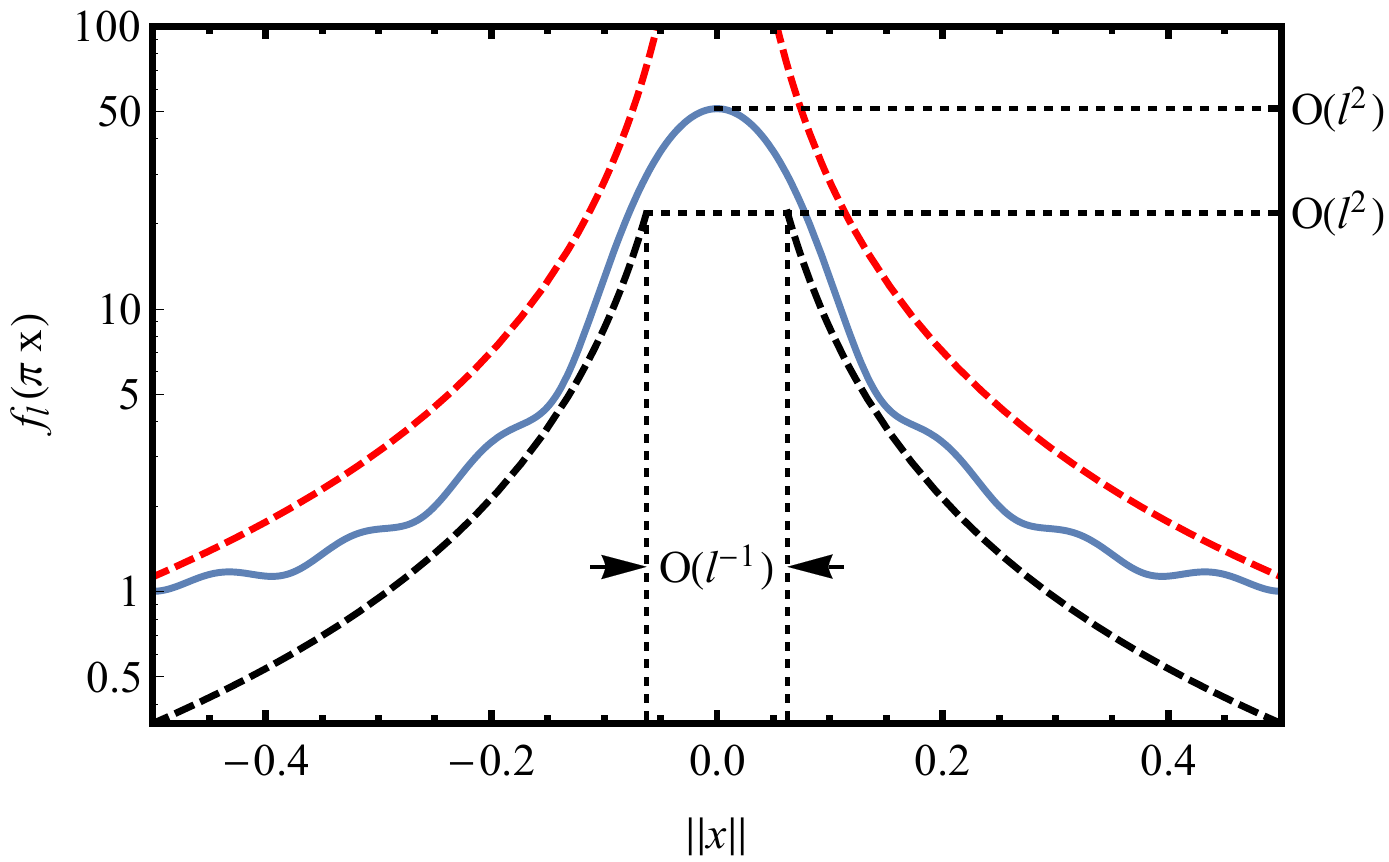}
\caption{
\emph{Elementary bounds on $f_\ell(\pi x)$:} The function $f_\ell(\pi x)$~\eqref{eq:fl} versus $\fracp{x}$ (blue, solid) is shown on a log-scale for $\ell=8$. The bounds~\eqref{eq:fbound} (red, dashed) and~\eqref{eq:flower} (black, dashed) are shown.}
\label{fig:fl}
\end{figure}

In this subsection we show that 
\begin{equation}
\tilde{C_\ell} = O(\log \ell).
\label{eq:Clesssim}
\end{equation}
for badly approximable numbers $Q/2\pi$ with $C_\ell$ from~\eqref{eq:Ctwiddle}.
We begin by noting that $f_\ell(\pi x)$ has maxima at integer $x$. As we approach the maxima the function grows as $\fracp{x}^{-2}$ before the divergence is cut off at $f_\ell(\mathbb{N})=(1+\ell)(1+2\ell)/3$, see Fig.~\ref{fig:fl}. It is thus straightforward to bound $f_\ell(\pi x)$ from above
\begin{equation}
f_\ell(\pi x) \leq \left(1+\frac{1}{\ell}\right)\frac{1}{4\fracp{x}^2}
\label{eq:fbound}
\end{equation}
(red in Fig.~\ref{fig:fl}). 
Equality is obtained if and only if $x$ is half-integer and $\ell$ is odd-integer.
We have from~\eqref{eq:Ctwiddle} and~\eqref{eq:fbound} that
\begin{align}
\tilde{C}_\ell &< \left(1+\frac{1}{\ell}\right) \sum_{k=1}^{\ell} \frac{1}{4k^2\fracp{k \zeta}^2} \nonumber \\
&\sim \int_{-\infty}^{\infty} \frac{M_\ell(x)\, dx}{4 x^2}
\label{eq:Cbar}
\end{align}
where 
\begin{align}
M_\ell(x) dx & \equiv \#\left\{k: k \fracp{k \zeta} \in (x, x+dx), 1\le k \le \ell \right\}
\end{align}
is the measure on the values of $x = k \fracp{k\zeta}$.

For almost all irrational numbers $\zeta$, the fractional part of $k \zeta$ becomes uniformly distributed in the interval $(-1/2,1/2)$ in the sense that
\begin{equation}
\#\left\{ k : \fracp{k \zeta} \in (y,y+\mathrm{d}y), 1\leq k \leq \ell   \right\} \sim \ell  \mathrm{d} y
\label{eq:evenmeasure}
\end{equation}
for large $\ell$. 
This can be shown using Weyl's criterion for the equidistribution of a sequence~\cite{finch2003mathematical}. 
It turns out that the density $M_\ell(x)$ for $x = k \fracp{k \zeta}$ scales as~\cite{harman1998metric}
\begin{align}
M_\ell (x) \mathrm{d}x & \sim \log \frac{\ell}{2|x|} \, \mathrm{d} x, \quad x_0 < |x| < \frac{\ell}{2}
\label{eq:logmeasure}
\end{align}
in the same limit.
This follows from Eq.~\eqref{eq:evenmeasure} under the stronger  condition that $\fracp{k\zeta}$ is uniformly distributed independent of $k$. 
In this case,
\begin{align}
M_\ell(x) &= \sum_{k=1}^\ell \delta(k \fracp{k \xi}-x) \nonumber\\
&\sim \sum_{k=1}^\ell \frac{1}{k} \theta(k/2-|x|) \nonumber\\
&\sim
\log \frac{\ell}{2|x|}, \, \, x_0 < |x|<\ell/2 
\end{align}
where in the last line we have assumed that $x>x_0$ from \eqref{eq:BadlyApprox}. Using this result in the RHS of Eq.~\eqref{eq:Cbar}:
\begin{equation}
\int_{-\infty}^{\infty}  \frac{M_\ell(x) dx}{4x^2} \sim \frac{\log\ell}{2 x_0}
\end{equation}
Hence the upper bound \eqref{eq:Clesssim} is shown. 

\subsection{Asymptotic lower bound on $\tilde{C}_\ell$}
\label{sec:Cgtrsim}

In this subsection we show that $\tilde{C_\ell}$ 
\begin{equation}
\tilde{C_\ell} = \Omega(\log \ell).
\label{eq:Cgtrsim}
\end{equation}
All the terms in~\eqref{eq:Ctwiddle} are positive, so we can lower bound $\tilde{C}_\ell$ by considering the sub-series of terms $k \leq \ell$ for which $k$ is equal to the denominator $q$ of a convergent $p/q$ of the irrational $\zeta$. The convergents $p/q$, where $p,q$ are co-prime, are the `best' rational approximations to $\zeta$ in the sense that
\begin{equation}
\left|\zeta - \frac{p}{q}\right| < \left|\zeta - \frac{p'}{q'}\right|
\label{eq:dioph}
\end{equation}
for all integers $p',q'$ such that $q'<q$. It is well known that the denominators of the convergents satisfy
\begin{equation}
q\fracp{q \zeta}<1.
\label{eq:Hurwitz}
\end{equation}
We use the elementary bound
\begin{equation}
f_\ell(\pi x) > \frac{1}{\pi^2 \fracp{x}^2}\left(1 - \frac{1}{2 \pi} \right) \, \mathrm{for} \fracp{x}>\frac{1}{2 \ell},
\label{eq:flower}
\end{equation}
shown in Fig.~\ref{fig:fl}. We can bound the relevant terms of the sum~\eqref{eq:Ctwiddle} by a constant using~\eqref{eq:Hurwitz}
\begin{equation}
\frac{f_\ell(\pi q \zeta)}{q^2} > \frac{1}{\pi^2}\left(1 - \frac{1}{2 \pi} \right) \mathrm{for} \fracp{q \zeta}>\frac{1}{2 	\ell}.
\label{eq:fC}
\end{equation}
The sum of this sub-series provides a bound on $\tilde{C}_\ell$
\begin{equation}
\tilde{C}_\ell > \sum_{q } \frac{f_\ell(\pi q \zeta)}{q^2}
\end{equation}
where the sum is over $q$ a denominator of a convergent as defined in~\eqref{eq:dioph} such that $\fracp{q \zeta} > 1/(2 \ell) $ and $q<\ell$. 

For badly approximable $\zeta$ there exists a constant $x_0>0$ such that $\fracp{q \zeta}>x_0/q$~(see \eqref{eq:BadlyApprox}). Thus $x_0/q > 1/(2\ell)$ implies $\fracp{q \zeta} > 1/(2 \ell)$. As the sum only includes $q<\ell$ we thus sum over the convergents satisfying the lower of these two bounds: $q<\ell \min(1,2x_0)$. Thus using~\eqref{eq:fC} to lower bound~\eqref{eq:Ctwiddle} we obtain
\begin{equation}
\tilde{C}_\ell > \frac{1}{\pi^2}\left(1 - \frac{1}{2 \pi} \right) \#\{q : q < \ell \min(1,2 x_0)\}.
\end{equation}
For badly approximable numbers (and furthermore algebraic and typical irrational numbers), the denominators $q$ of the convergents are exponentially spaced. Thus below a bound $q \leq q_{\max} \sim c \ell$ there are a number of $q$ that scales as $\# \{ q: q< q_{\max} \} \sim c' \log \ell$, and hence~\eqref{eq:Cgtrsim} follows.

\subsection{Constant bound on a summation of terms $k>\ell$ from the Ces\`{a}ro sum $C_\ell$}
\label{App:bound}

Here we show that $C_\ell'= O (\ell^0)$ and $\sigma_\ell'  = O (\ell^0)$ where 
\begin{equation}
C_\ell' := \sum_{k = \ell+1}^{\infty} \frac{f_\ell(Q k/2)}{k^2} 
\end{equation}
with $f_\ell$ from~\eqref{eq:fl} and 
\begin{equation}
\sigma_\ell' := \sum_{k = \ell+1}^{\infty} \frac{\sin^2(Q k \ell /2)}{k^2 \sin^2(Q k /2)}. 
\end{equation}
This result is used in Appendices~\ref{App:LogWander} and~\ref{app:sigmabound}. We split the sum into blocks of $m$ terms and re-weight the terms to obtain an upper bound 
\begin{equation}
\begin{aligned}
C_\ell' & = \sum_{n = 0}^\infty \sum_{k = \ell+n m+1}^{\ell + (n+1)m} \frac{f_\ell(Q k/2)}{k^2}
\\
& < \sum_{n = 0}^\infty  \frac{1}{(\ell+n m +1)^{2}}\sum_{k = \ell+n m+1}^{\ell + n m +m} f_\ell(Q k/2)
\end{aligned}
\end{equation}
As $f_\ell(\theta)$ is a smooth, bounded $\pi$-periodic function, and $Q/2\pi$ is irrational, the sum converges on an integral
\begin{equation}
\begin{aligned}
\sum_{k = \ell+n m+1}^{\ell + n m +m} f_\ell(Q k/2) & = m(\ell+1)(1+ O(m^{-1}))\\
& =  m(\ell+1) + O(\ell^1 m^0) .
\end{aligned}
\end{equation}
where the sub-leading terms are the usual error from approximating a sum with an integral, and we have used that
\begin{equation}
\frac{1}{2 \pi} \int_{-\pi}^\pi \mathrm{d}\theta f_\ell(\theta) = \ell+1.
\end{equation}
Replacing the sum with an integral is permitted by the equidistribution of the numbers $Q k/2 \mod 2 \pi$, which holds for all irrational numbers~\cite{finch2003mathematical}. Thus we have a bound
\begin{equation}
C_\ell' <  \left( m(\ell+1) + \mathrm{O}(\ell^1 m^0) \right) \sum_{n=0}^{\infty} \frac{1}{(\ell+n m +1)^{2}}.
\label{eq:AppCPrime}
\end{equation}
In order to simplify the sum, observe that for a monotonically decreasing function $g(x)$ 
\begin{equation}
\sum_{n=0}^\infty g(n) < g(0) + \int_0^\infty \mathrm{d}x \,g(x),
\end{equation}
which follows from comparing the terms $n = 1, \ldots \infty$ with the left Reimann sum. Using this result to bound the sum in~\eqref{eq:AppCPrime} we obtain
\begin{equation}
\begin{aligned}
C_\ell' &< \left( m(\ell+1) + O(\ell^1 m^0) \right) \left( \frac{1}{(\ell+1)^2} + \frac{1}{m(\ell+1)} \right)
\\
&= 1 + \frac{m}{(\ell+1)} + O(\ell^{0} m^{-1}) + O(\ell^{-1} m^{0})
\end{aligned}
\end{equation}
Thus if we choose our initial separation of the sum into blocks of size $m$ such that $m$ scales as $m \sim \ell^\alpha$ with $0< \alpha \leq 1$, then $C_\ell'$ is bounded by a constant. 

The same argument is applied to $\sigma_\ell'$ by making the replacement
\begin{equation}
f_\ell(\theta) \rightarrow g_\ell(\theta):= \frac{\sin^2(\ell \theta)}{\sin^2(\theta)}
\end{equation}
which satisfies
\begin{equation}
\frac{1}{2 \pi} \int_{-\pi}^{\pi} \mathrm{d} \theta g_\ell(\theta) = \ell.
\end{equation}
By following exactly the same procedure and splitting the sum into blocks of $m$ we obtain the similar bound
\begin{equation}
\begin{aligned}
\sigma_\ell' &:= \sum_{k = \ell+1}^{\infty} \frac{g_\ell(Q k/2)}{k^2} \\
& < \left( m \ell + O(\ell^1 m^0) \right) \left( \frac{1}{(\ell+1)^2} + \frac{1}{m(\ell+1)} \right)
\end{aligned}
\end{equation}
thus 
\begin{equation}
\sigma_\ell' = O(\ell^0)
\label{eq:sigellbound}
\end{equation}
which holds, as before, for an appropriate choice of scaling between $m$ and $\ell$.

\subsection{Asymptotic upper bound on $\sigma^2(S_\ell)$}
\label{app:sigmabound}

In this section we use results derived in the analysis of $C_\ell$ to show that
\begin{equation}
\sigma^2(S_\ell) = O(\log \ell).
\label{eq:sigmabound}
\end{equation}
for strong smooth modulation, when $|\delta_k|=O(k^{-2})$~\eqref{eq:AppDelkFinal}.
Using the Fourier representation of $\sigma^2(S_\ell)$ from~\eqref{eq:SigEll},  and the elementary bounds 
\begin{align}
\sin^2(Q k \ell/2) & = \sin^2(\pi k \zeta \ell) = O(1) \\
\frac{1}{\sin^2(Q k /2)} &= \frac{1}{\sin^{2}(\pi k \zeta)} = O\left(\frac{1}{\fracp{k \zeta}^{2}}\right)
\end{align}
one obtains
\begin{equation}
\sigma^2(S_\ell) = O \left( \sum_{k=1}^\infty \frac{1}{k^2 \fracp{k \zeta}^2} \right).
\end{equation}
Splitting this sum into terms $k>\ell$ and $k \leq \ell$. The terms $k>\ell$ are bounded by a constant~\eqref{eq:sigellbound} whereas the terms $k \leq \ell$ grow logarithmically with $\ell$~\eqref{eq:Clesssim}. This proves the proposition~\eqref{eq:sigmabound}.

\section{Exact logarithmic wandering coefficient for $Q/2 \pi = \tau$, the Golden ratio}
\label{App:wGolden}

In this section we show
\begin{equation}
w = \lim_{\ell \to \infty} \frac{\tilde{C}_\ell}{\log \ell} = \frac{2\pi^2}{15\sqrt{5} \log \tau}
\label{eq:AppW}
\end{equation}
where $Q/2\pi = \tau = (1+\sqrt{5})/2$ and, as before,
\begin{equation}
\tilde{C}_\ell = \frac{1+2\ell}{2 \ell} \sum_{k=1}^\ell \frac{1}{k^2\sin^2(\pi k\tau)} \left(1 - \frac{\sin((1+2 \ell)\pi k \tau)}{(1+2 \ell)\sin (\pi k \tau)} \right).
\end{equation}
The first simplification comes from observing that the second term in the brackets above, given by
\begin{equation}
\frac{1+2\ell}{2 \ell} \sum_{k=1}^\ell \frac{1}{k^2\sin^2(\pi k\tau)} \left( \frac{\sin((1+2 \ell)\pi k \tau)}{(1+2 \ell)\sin (\pi k \tau)} \right),
\end{equation}
is bounded between its finite extremal values realised for $\ell = 4$ and $\ell = 8$. Thus the asymptotic scaling is given by the leading term only,
\begin{equation}
w = \lim_{\ell \to \infty} \frac{1}{\log \ell}  \sum_{k=1}^\ell \frac{1}{k^2\sin^2(\pi k\tau)}.
\end{equation}
This limit happens to be a known result, evaluated by D. Speyer in Ref~\cite{speyerprivate} yielding the result~\eqref{eq:AppW}.

\section{$\hat\delta_k$ for sinusoidal modulation}
\label{app:dk}

In this Appendix we evaluate the asymptotic (large $k$) form of the Fourier transformed reduced coupling
\begin{equation}
\hat\delta_k = \frac{1}{2 \pi} \int_{-\pi}^\pi \mathrm{d} \theta \mathrm{e}^{i k \theta} \log \left| \frac{\J(\theta)}{\h(\theta)} \right|
\end{equation}
when $\J(\theta)$, $\h(\theta)$ are smooth periodic functions. 

For weak modulation $\delta(\theta)$ is also a smooth periodic function and hence $\delta_k$ decays exponentially. 

At strong modulation the zeros of $\J(\theta)$, $\h(\theta)$ induce logarithmic divergences in $\delta(\theta)$. These singularities control the large $k$ behaviour of $\delta_k$:
\begin{equation}
\begin{aligned}
\frac{1}{2 \pi} \int_{-\pi}^\pi \mathrm{d} \theta \mathrm{e}^{i k \theta} \log \left| \J(\theta) \right| & \sim \frac{1}{2 \pi} \sum_{\theta_0} \int_{-\infty}^{\infty} \mathrm{d} \theta \mathrm{e}^{i k \theta} \log |\theta-\theta_0| 
\\
& \sim - \sum_{\theta_0} \mathrm{e}^{i k \theta_0} \frac{1}{2 |k|}
\end{aligned}
\end{equation}
were $\theta_0$ are the solutions to $\J(\theta_0)=0$.

The independence of the second line on any of the data of $\J(\theta)$ other than the positions of its roots is consistent with the observed stability of the critical properties of the QP Ising transition to tuning of the parameters $\J,\h,\AJ,\Ah$. 
Using the equivalent result for the contribution from $\h(\theta)$ we obtain,  
\begin{equation}
\hat\delta_k \sim \frac{1}{2 |k|} \left( 
\sum_{\theta_0:\h(\theta_0)=0} \mathrm{e}^{i k \theta_0}
-
\sum_{\theta_0:\J(\theta_0)=0} \mathrm{e}^{i k \theta_0}
\right).
\label{eq:AppDelk}
\end{equation}
With this we obtain
\begin{equation}
|\hat\delta_k|^2 = O( k^{-2})
\label{eq:AppDelkFinal}
\end{equation}
and
\begin{equation}
\cexp{|\delta_k|^2 k^{2}}_\textrm{Ces\'{a}ro} = \frac{1}{k}\sum_{k'=1}^{k} |\delta_{k'}|^2 {k'}^{2} \sim 1.
\label{eq:dkcesaro}
\end{equation}
where we have assumed that all of the roots are first order (i.e. that $|\frac{\mathrm{d} J}{\mathrm{d} \theta}|_{\theta = \theta_0}>0$ and similarly for $\Gamma$).

For rare special values of the positions of the zeros $\theta_0$ the asymptotic form of $|\hat\delta_k|^2$ is enhanced by a constant factor. As these points are fine-tuned we do not discuss this further. 

\section{Estimate of $z$}
\label{App:zEstimate}
Here we estimate $z$ by evaluating  the maximum energy $\epsilon_0(q)$ of the lowest band of excitations in the commensurately modulated TFIM with period $q$. As is shown in Figure~\eqref{fig:Z}, the QP-Ising critical point exhibits single parameter scaling such that as $q \to \infty$, $\epsilon_0(k,q) \sim c q^{-z} $ where $z \approx 1.9$ is the dynamical exponent.

The energies of the Fermionic modes with crystal momentum $k$ are given by the roots of the characteristic equation 
\begin{equation}
\det(H(k) - \epsilon) = \sum_{k=0}^q \chi_{2k} \epsilon^{2k}.
\end{equation}
where $H(k)$ is the Hermitian $2q\times 2q$ cyclic-tri-diagonal matrix 
\begin{equation}
H(k) = \frac{i}{4}
\begin{pmatrix}
 & \h_1 & & & & \cdots & -\J_q \mathrm{e}^{i k q} \\
 -\h_1 & & \J_1 & & & &  \\
 & -\J_1 & & \h_2 & & & \\
 & & -\h_2 & & \J_2 & & \\
 & & & -\J_2 & & \ddots & \vdots\\
 \vdots & & & & \ddots & & \h_q \\
 \J_q \mathrm{e}^{-i k q} &  & & &\cdots & -\h_q &  \\
\end{pmatrix}.
\label{eq:Hk}
\end{equation}
This characteristic equation is studied in more detail in Refs~\cite{han1994critical,Chandran:2017ab}. The eigen-energies of this Hamiltonian come in $\pm\epsilon$ pairs; the positive eigen-energies correspond to the excitation spectra of the TFIM. We denote the smallest positive eigen-energy as $\epsilon_0$. Here we use the following two properties of the Hamiltonian $H(k)$ when at QP-Ising criticality: (i) the coefficient $\chi_0$ is given by 
\begin{equation}
\chi_0 = (-1)^q 2 (1 + \sigma \cos k q) \prod_{i=1}^{q} |\J_i \h_i| 
\label{eq:chi0}
\end{equation}
where $\sigma = \textrm{sign}(\prod_i \J_i \h_i)$. (ii) the coefficient of $\epsilon^2$, is given by
\begin{equation}
\chi_{2} = q (-1)^{q-1} \left( \prod_i |\J_i \h_i| \right) \sum_{\ell=0}^{q-1} \cexp{\frac{\mathrm{e}^{2 S_\ell(i)}}{|\h_{i+\ell}|^2}}_{i}
\label{eq:Chi2} 
\end{equation}
where $\cexp{\cdot}_i$ indicates spatial averaging of the index $i = 1,2 \ldots q$ and $S_\ell(i)$ is defined in~\eqref{eq:SL}. In order to estimate the value $z$ we truncate the characteristic equation to quadratic order
\begin{equation}
\chi_2 \epsilon_0^2 + \chi_0 = 0.
\end{equation}
From~\eqref{eq:chi0} this puts the extrema of the lowest band at $k q = 0, \pi$. We consider the scaling of the greater of the two energies by taking $\sigma \cos(k q)=1$.

We obtain an estimate for $\epsilon_0$ by (i) neglecting the correlations between $S_\ell(i)$ and $\h_{i+\ell}$, (ii) replacing the factor $\h_{i+\ell}$ with a single characteristic energy scale $\bar{\h}$, and (iii) treating the $S_\ell(i)$ as Gaussian independently distributed variables with mean $\mu = 0$ and variance $\sigma(S_\ell(i))^2 = w \log \ell$. This neglects both correlations between $S_\ell(i)$ for different $i$, and non-Gaussian statistics in the higher order moments of the distribution of each $S_\ell(i)$. We can then use that for Gaussian distributed random variables with mean $\mu$ and variance $\sigma^2$
\begin{equation}
\cexp{\mathrm{e}^{a x}}_x =  \mathrm{e}^{a \mu + a^2 \sigma^2/2}.
\end{equation}
Making this approximation yields
\begin{align}
-\frac{\chi_2}{\chi_0} &= \frac{q}{4} \sum_{\ell=0}^{q-1} \cexp{\frac{\mathrm{e}^{2 S_\ell(i)}}{|\h_{i+\ell}|^2}}_{i} \nonumber \\
& \approx \frac{q}{4 \Bh^2} \sum_{\ell=0}^{q-1}  \cexp{\mathrm{e}^{2 S_\ell(i)} }_{i} 
\nonumber \\
& \approx \frac{q}{4 \Bh^2} \sum_{\ell=0}^{q-1}  \ell^{2 w} 
\nonumber \\
& 
\sim \frac{q^{2+2 w}}{4(1+2 w)\Bh^2}.
\end{align}
This gives an estimated typical scaling of the lowest bandwidth
\begin{equation}
\epsilon_0 = \sqrt{-\frac{\chi_0}{\chi_2}} \sim c q^{-1-w}
\end{equation}
yielding $z \approx 1+w$. For $Q/2 \pi = \tau$ the Golden ratio, we find (App.~\ref{App:wGolden}) that
\begin{equation}
z \approx 1 + w = 1 + \frac{2 \pi^2}{15 \sqrt{5} \log \tau} = 2.22\ldots.
\end{equation}
which is approximately 15\% larger than the value $z\approx 1.9$ identified numerically in the main text.

\bibliography{paper-bib}

\end{document}